\documentclass[a4paper,twocolumn,11pt]{quantumarticle}

\pdfoutput=1

\usepackage[utf8]{inputenc}
\usepackage[english]{babel}
\usepackage[T1]{fontenc}
\usepackage{amsmath,amssymb,amsfonts,amsthm,mathtools}
\usepackage{bm}
\usepackage{dsfont}
\usepackage{mathbbol}
\usepackage{graphicx}
\usepackage[numbers]{natbib}
\usepackage{hyperref}
\hypersetup{hidelinks}

\newcommand{\idop}{\mathbbm 1}

\newcommand{\Trace}{{\rm Tr}}

\newcommand{\re}{{\rm Re}}

\def \ket#1{\mathinner{|{#1}\rangle}}
\def \bra#1{\mathinner{\langle{#1}|}}

\def\braket#1{\mathinner{\langle{#1}\rangle}}

\newcommand{\ketbra}[2]{{\mathinner{| {#1} \rangle \langle {#2} |}} }

\newcommand{\ha}{\hat{a}}
\newcommand{\hA}{\hat{A}}
\newcommand{\hu}{\hat{u}}
\newcommand{\hU}{\hat{U}}

\newcommand{\mathG}{\mathcal{G}}

\begin{document}

\title{Quantum random-number generator with non-demolition measurements: semi-device-independent implementation}

\author{Paolo Solinas}
%\email{paolo.solinas@unige.it}
\email{paolo.solinas@ge.infn.it}
\affiliation{Dipartimento di Fisica, Universit\`a di Genova, via Dodecaneso 33, I-16146, Genova, Italy.}
\affiliation{INFN - Sezione di Genova, via Dodecaneso 33, I-16146, Genova, Italy.}
\author{Giovanni Chesi}
\affiliation{QUIT Group, Physics Department, University of Pavia, via Bassi
6, Pavia, 27100, Italy}
\affiliation{INFN Sez. Pavia, via Bassi 6, Pavia, 27100, Italy}

%\date{\today}

%\pacs{85.80.Lp,74.50.+r,72.25.-b}
\begin{abstract}
We propose and analyze a novel quantum random-number generator based on a tripartite quantum system in which two subsystems act as detectors. Within a quantum non-demolition measurement scheme, one detector is used to certify the presence of genuine quantum effects in the system's evolution, while the second generates random numbers from a distribution that can be optimized to maximize their entropy. Using one two-level system and two three-level systems, we generate random numbers from a nearly uniform three-outcome distribution, yielding close-to-maximal entropy and therefore near-optimal randomness generation.
A key feature of the protocol is that randomness generation and certification occur simultaneously. Moreover, certification does not rely on spacelike separation between detectors, removing a major constraint of device-independent approaches. This property enables practical implementation and facilitates the miniaturization of the device, making the protocol a promising candidate for scalable quantum technologies.
\end{abstract}

\maketitle

%%%%%%%%%%%%%%%%%%%%%%%%%%%%%%%%%%%%%%%%%%%%%%%%%%%%%%%%%%%%
\section{Introduction}
In quantum mechanics, the measurement process is intrinsically probabilistic. While this often implies that numerous experimental runs are required to characterize the state of a system, it can also be exploited to perform efficient and secure tasks such as quantum cryptography \cite{bb84,nielsen-chuang_book} and random-number generation \cite{colbeck2009quantum,Pironio2010}.

The use of quantum devices for random-number generation overcomes several limitations that typically affect classical generators, such as finite periodicity and vulnerability to third-party attacks. However, for practical technological applications, it is essential that the end user does not need to understand quantum mechanics or the internal functioning of the device. At the same time, the user must be able to certify that the device genuinely exploits quantum phenomena and that the generated random numbers are intrinsically random. This requirement underlies the concept of device-independent random-number generation \cite{colbeck2009quantum,Pironio2010,acin2007}.

The original proposal for a quantum random-number generator (QRNG) exploited nonlocality and entanglement \cite{colbeck2009quantum,Pironio2010}. In a Bell-like setup \cite{Bell1964}, a string of random numbers is extracted from two spacelike-separated devices, and a Bell inequality test is subsequently performed on the collected data \cite{colbeck2009quantum,Pironio2010,e91}. If the Bell inequality is violated, one can certify that the random numbers are generated by a quantum process and are therefore intrinsically random.

This idea has been further developed and refined over the past decades \cite{Brunner2014, Cavalcanti2012, Bierhorst2018, Li2021, Liu2021, Liu2025}. Despite these important technical improvements, it faces an undesirable yet physically unavoidable limitation: to certify a violation of Bell inequalities, the devices must be spatially separated. This requirement imposes a significant constraint on technological applications.

A possible solution is to certify the quantumness of the generated data by exploiting other quantum features, such as coherent superposition or the invasiveness of the measurement process. In this direction, Leggett and Garg proposed in 1985 an alternative framework based on a temporal sequence of measurements of non-commuting observables \cite{Leggett1985}. They derived an inequality that is formally analogous to Bell’s inequality. If the collected data violate the Leggett–Garg inequality (LGI), the system can be certified as quantum. The idea of using LGIs for QRNGs was originally proposed and implemented in Refs.~\cite{Mal2016,Nath2024}.

The main advantage of this approach is that sequential measurements are performed on a single system, enabling device miniaturization. Existing implementations ~\cite{Mal2016,Nath2024}, however, offer limited parameter flexibility and relatively low generation rates.

Here, we propose a change of paradigm that overcomes these limitations through a new QRNG based on the quantum Non-Demolition Measurement (QNDM) protocol \cite{solinas2015fulldistribution, solinas2016probing, solinas2021, solinas2022, solinas2025, solinasAdvPhysX2026}. Indeed, it has been shown that QNDM provides a stronger criterion than the LGI for the identification and certification of quantum behavior \cite{solinas2025, solinasAdvPhysX2026}.

The protocol employs a quantum system coupled to two ancillary detectors, both measured at the end of the evolution. The first detector
 certifies path interference within the QNDM framework, whereas the second one generates random outcomes associated with the system’s evolution paths. Generation and certification therefore occur simultaneously but are extracted from distinct detectors, rather than certification being performed \textit{a posteriori} on the generated data \cite{Pironio2010,Nath2024}.

%%%%%%%%%%%%%%%%%%%%%%%%%%%%%%%%%%%%%%%%%%%%%%%%%%%%%%%%%%%%
\section{System and set-up}
\begin{figure}
    \begin{center}
    \includegraphics[scale=.5]{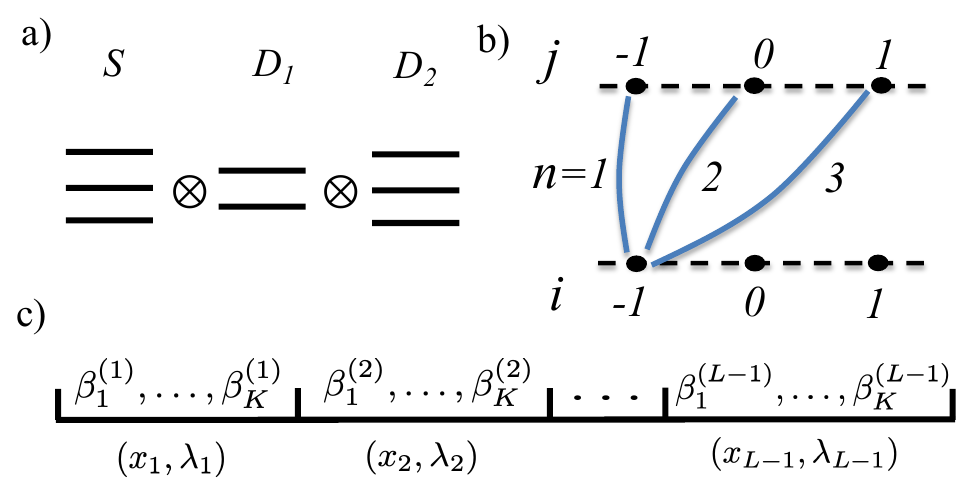}
    \end{center}
    \caption{a) The spectrum of the system $S$ and ancillary detectors $D_1$ and $D_2$ with three, two, and three quantum states.
    b) Three representative paths out of nine traversed by the system $S$. The paths $\{a_i, a_j \} = \{-1,-1\} \rightarrow n=1$, $\{-1,0\}\rightarrow n=2$ and $\{-1,1\}\rightarrow n=3$ are shown.
    c) The random string extracted from $D_2$. For a fixed $\lambda_l$, we obtain a string $x_l$ that is composed of $K$ integers $\beta_n^{(l)} = 0, \pm 2$ extracted from the distribution \eqref{eq:Prob_D2}. 
    }
    \label{fig:sup_path}
\end{figure}

We consider a tripartite system composed of a three-level system $S$, a two-level detector $D_1$, and a three-level detector $D_2$, as shown in Fig.~\ref{fig:sup_path} a).

As in non-demolition protocols \cite{solinas2015fulldistribution, solinas2025, solinasAdvPhysX2026}, the information about the observables is encoded in the phase of the detectors. 
In the full space, the information encoding operation corresponds to coupling $S$ to $D_1$ through the unitary operator $\hat{U}_1=\exp\{ i \hat a \otimes \hat\lambda \}$, and $S$ to $D_2$ through $\hat{U}_2=\exp\{ \frac{2 i \pi}{M} \hat A \otimes \hat m \}$.
The operators $\hat a$ and $\hat A$ act on the degrees of freedom of $S$, while $\hat\lambda$ and $\hat m$ act on $D_1$ and $D_2$ and satisfy the eigenvalue equations $\hat\lambda \ket{\lambda} = \lambda \ket{\lambda}$ and 
$\hat m \ket{m} = m \ket{m}$, respectively.

In the following, we assume that $\hat a$ and $\hat A$ commute and therefore share a common eigenbasis. To simplify the notation, we label the common eigenstates with a single index, i.e., $\ket{i} \equiv \ket{a_i, A_i}$, such that $\hat a \ket{i} = a_i \ket{i}$ and $\hat A \ket{i} = A_i \ket{i}$. Since $S$ is a three-dimensional system, without loss of generality, we take $i = 1, 0, -1$. These states correspond to the eigenvalues $a_i = i$, i.e., $a_1 = 1$, $a_0 = 0$, and $a_{-1} = -1$, and to $A_i = 1$ for $i = 1, 0$, and $A_{-1} = -1$. Thus, $\hat a$ is nondegenerate, whereas $\hat A$ is doubly degenerate.

%%%%%%%%%%%%%%%%%%%%%%%%%%%%%%%%%%%%%%%%%%%%
%%%%%%%%%%%%%%%%%%%%%%%%%%%%%%%%%%%%%%%%%%%%

With this notation, the initial states of the three subsystems are taken to be $\ket{\psi_0}=\sum_{i} \psi_i^0 \ket{i}$ for $S$, $1/\sqrt{N} \sum_{\lambda}  \ket{\lambda}$ for $D_1$, and $1/\sqrt{M} \sum_{m}  \ket{m}$ for $D_2$. The corresponding density matrices are denoted by $\rho_0$, $\rho_{0, D_1}$ and $\rho_{0, D_2}$, respectively.

The total unitary operator for the two sequential non-demolition measurements reads
$\hU = \hU_1 \hU_2 \hu \hU_1 \hU_2$ where $\hu$ is a generic unitary transformation acting on $S$.

The evolution of the total system for the sequential couplings leads to the density matrix (see Appendix for more details)
\begin{eqnarray}
    \rho_f &=& \frac{1}{\mathcal{N}} \sum 
    u_{j,i} \rho^0_{i,i'} u^\dagger_{i',j'}
    e^{i \left[ \lambda (a_i+a_j)- \lambda' (a_{i'}+a_{j'}) \right]} \nonumber \\
    &&    
    e^{i \frac{2 \pi}{M}
    \left[m (A_i+A_j) - m' (A_{i'}+A_{j'})\right] } \ketbra{j, \lambda, m}{j', \lambda', m'} \nonumber \\
    \label{eq:rho_tot_fin}
\end{eqnarray}
where $\mathcal{N} = NM$, $u_{j,i} = \bra{j} \hat{u} \ket{i}$, $\rho^0_{i,i'} = \bra{i} \rho_0 \ket{i'}$, and the sum is over the indices $\{ i,j,\lambda,m, i', j', \lambda', m'\}$.

It is convenient to adopt an alternative notation.
A path in the eigenstate configuration space is associated with an ordered pair $\{i, j\}$.
Since we have a three-state system, we have a total of $3^2=9$ paths, so the path index is $1 \leq n \leq 9$ (see Fig. \ref{fig:sup_path} b).
A possible mapping between the pair $\{ i, j \}$ and an integer $n$ is $n= 1+ 3 (i+1) +( j+1)$; so that, for example, $\{ i, j \} = \{ 1, 1\} \rightarrow n = 9$ and $\{ i, j \} = \{-1, -1\} \rightarrow n = 1$. 

Since the final measurements occur on the detectors, we trace out the degrees of freedom of $S$, corresponding to taking $j'= j$ in Eq. \eqref{eq:rho_tot_fin} (see Appendix).
With the notation change, the detectors' density matrix becomes
\begin{eqnarray}
    \bar{\rho}_f &=& \frac{1}{\mathcal{N}} %\sum_{k,n=1}^9 
    \sum
    \mathcal{A}_{n,k} 
    e^{i \left( \lambda \alpha_n- \lambda' \alpha_k \right)} 
    %e^{i \frac{ \pi}{M}  \left(m\beta_n - m' \beta_k\right) }
    e^{i \frac{2 \pi}{M}  \left(m\beta_n - m' \beta_k\right) } \nonumber \\
    && \ketbra{ \lambda, m}{ \lambda', m'}
    \label{eq:rho_D_path}
\end{eqnarray}
where the sum is over the indices $\{ k, n,\lambda, \lambda', m, m'\}$, $\mathcal{A}_{n,k}=u_{j,i} \rho^0_{i,i'} u^\dagger_{i',j}$ and $\alpha_n$, $\alpha_k$, $\beta_n$, $\beta_k$ are associated by the above mapping to $a_i+a_j$, $a_{i'}+a_j$, $A_i+A_j$, $A_{i'}+A_j$, respectively.
In practice, we perform the calculation with the path notation, but then, to have explicit values, we must revert to the mapping with the original $\{a_i,a_j\}$ notation.
Notice that, with the path notation, since the final system state is fixed to $j$, the combinations $\{n,k\}$ with different $j$ are not allowed in Eq. \eqref{eq:rho_D_path} (see Appendix).

%%%%%%%%%%%%%%%%%%%%%%%%%%%%%%%%%%%%%%%%%%%%%%%%%%%%%%%%%%%%
For the quantumness certification, we focus on $D_1$, and in Eq.~\eqref{eq:rho_D_path}, we trace out the $D_2$ degrees of freedom \cite{solinas2015fulldistribution, solinasAdvPhysX2026}, obtaining the state $\rho_{f,D_1} \equiv \Trace_{D_2}[\bar{\rho}_f]$. By interferometric techniques, we measure the accumulated phase between the states $\ket{\pm \lambda}$ 
and we obtain the quasi-characteristic function $\mathG_\lambda = \braket{\lambda | \rho_{f, D_1}|-\lambda}/\braket{\lambda | \rho_{0, D_1} |-\lambda}$; that is,
%$\mathG_\lambda = \braket{\lambda | \rho_{f, D_1}|-\lambda}/\braket{\lambda | \rho_{0, D_1} |-\lambda} = \sum_{k,n=1}^9  f(n,k)     \mathcal{A}_{n,k}     \exp\{i  \lambda \left(\alpha_n+ \alpha_k \right)\}$, 
\begin{equation}
    \mathG_\lambda  
    %= \sum_{k,n=1}^9 \left( \frac{1}{M} \sum_{m} e^{i \frac{2 \pi m}{M}  \left( \beta_n -  \beta_k\right) } \right)\mathcal{A}_{n,k} e^{i  \lambda \left(\alpha_n+ \alpha_k \right)} 
    = \sum_{k,n=1}^9 
    f(n,k)
     \mathcal{A}_{n,k} 
    e^{i  \lambda \left(\alpha_n+ \alpha_k \right)}.
    \label{eq:G_lambda}
\end{equation}
where 
$f(n,k) =1/M~ \sum_{m} \exp\{i \frac{2 \pi m}{M}  \left( \beta_n -  \beta_k\right) \}$ and the initial coherence of $D_1$ is $\braket{\lambda | \rho_{0, D_1}|-\lambda} = 1/N$.
The function $f(n,k)$ accounts for the entanglement with $D_2$ and acts as a decoherence functional for $D_1$.

Since $\alpha_n+ \alpha_k \rightarrow 2 a_j + a_i+a_{i'}$ and recalling that $a_j, a_i, a_{i'} = 0, \pm 1$ we have that $\alpha_n+ \alpha_k = 0, \pm 1, \pm 2, \pm 3, \pm 4$.
As a consequence, apart from the case with $\alpha_n+ \alpha_k =0$, the periodicity of $\mathG_\lambda$ in $\lambda$ is $2 \pi$, i.e., $\mathG_\lambda = \mathG_{\lambda + 2 \pi}$.

The corresponding quasi-probability distribution density (QPDD) is obtained by taking the inverse Fourier transform of $\mathcal{G}_\lambda$, namely,
$P_{ND} (\Delta) = 1/(2\pi) \int d \lambda \exp\{-i \lambda \Delta \}\mathG_\lambda = \sum_{k,n=1}^9 f(n,k)\mathcal{A}_{n,k}~\delta \left[\Delta - (\alpha_n+ \alpha_k) \right]$ \cite{solinas2025, solinasAdvPhysX2026}.

It is useful to decompose $P_{ND}(\Delta)$ into a classical and a quantum contribution,
$P_{ND} (\Delta) = P_{cl} (\Delta) + P_{q} (\Delta)$ where $P_{cl} (\Delta) = \sum_{n=1}^9 P_n~\delta \left[\Delta - 2\alpha_n \right]$ and $P_{q} (\Delta) = 2 \sum_{n>k} \re \left[f(n,k)
     \mathcal{A}_{n,k} \right] 
 \delta \left[\Delta - (\alpha_n+ \alpha_k) \right]$.
The contribution $P_n=\mathcal{A}_{n,n}$ is the probability associated with the $n$-th path.
As shown in Refs.~\cite{solinas2025,solinasAdvPhysX2026}, $P_{ND}(\Delta)$ is normalized, i.e., $\sum_{n=1}^{9} P_n=1$,
while the quantum contribution satisfies $2 \sum_{n>k}
\mathrm{Re}\!\left[f(n,k)\mathcal{A}_{n,k}\right]=0$.

The classical term $P_{cl}(\Delta)$ describes trajectories associated with definite paths in the eigenvalue space, e.g., $a_i\rightarrow a_j$, with the corresponding (positive) probability $P_n$.
By contrast, $P_q(\Delta)$ originates from the coherent superposition of different paths and, therefore, contains the interference contributions generated during the evolution.

Since the total quantum contribution integrates to zero and is real \cite{solinas2025,solinasAdvPhysX2026}, any nonvanishing $P_q(\Delta)$ must contain both positive and negative terms. Consequently, whenever the dynamics involves coherent superpositions of paths, the quasi-probability distribution necessarily develops negative regions \cite{solinas2025,solinasAdvPhysX2026}. Conversely, in the absence of path superpositions, $P_q(\Delta)=0$ and $P_{ND}(\Delta)$ reduces to a genuine probability distribution. Therefore, the negativity of $P_{ND}(\Delta)$ provides a necessary and sufficient signature of path superposition in the dynamics \cite{solinas2025,solinasAdvPhysX2026}. Such negative regions cannot be reproduced by a classical device that generates the corresponding strings from an underlying probability distribution.

%%%%%%%%%%%%%%%%%%%%%%%%%%%%%%%%%%%%%%%%%%%%%%%%%%%%%%%%

Detector $D_1$ is used for certification, whereas $D_2$ generates random numbers.  Starting from Eq.~\eqref{eq:rho_D_path}, we trace over the degrees of freedom of $D_1$ and obtain the reduced density matrix of the second detector. We then apply the Quantum Fourier Transform \cite{nielsen-chuang_book}, obtaining $\rho_{f, D_2}
= \sum_{k,n=1}^{9}
F(n,k)\,\mathcal{A}_{n,k}\,
\ketbra{\beta_n}{\beta_k}$,
where $F(n,k)=
\cos\!\left[\lambda(\alpha_n-\alpha_k)\right]$
(see Appendix).
Like $f(n,k)$ introduced above, $F(n,k)$ acts as a decoherence functional, partially suppressing the interference contributions associated with superposed paths. 

We recall that, since $\hat A$ has degenerate eigenvalues, different paths labelled by different values of $n$ can correspond to the same detector outcome. For example, the pairs $\{i,j\}=\{1,1\}$ and $\{i,j\}=\{1,0\}$ correspond to different paths, namely $n=9$ and $n=8$, respectively, but yield the same value of $\beta_n$, i.e., $\beta_9=\beta_8=2$. As a consequence, the possible outcomes of detector $D_2$ are $\beta_n=0,\pm 2$. 
This degeneracy is essential: paths associated with the same $\beta_n$ are indistinguishable to $D_2$, so their amplitudes can interfere and produce the signal measured by $D_1$, thereby enabling quantum certification.

The probability of obtaining a given outcome $\beta_n$ is therefore obtained by summing over all paths associated with the same value of $\beta_n$ (see Appendix):
\begin{equation}
    \mathcal{P}_n(\lambda)
    =
    \sum_{m}' P_m
    + 2
    \sum_{m>k}'
    \mathrm{Re}\!\left(\mathcal{A}_{m,k}\right)
    \cos\!\left[\lambda(\alpha_m-\alpha_k)\right],
    \label{eq:Prob_D2}
\end{equation}
where the primed sums run over all paths corresponding to the same degenerate value of $\beta_n$.

Equation~\eqref{eq:Prob_D2} summarizes the complementary roles of the two detectors. Measurements of $D_1$ reconstruct the quasicharacteristic function $\mathcal{G}_\lambda$, while each measurement of $D_2$ yields an outcome $\beta_n=0, \pm 2$ distributed according to $\mathcal{P}_n(\lambda)$. The dependence on $\lambda$ allows the output statistics to be controlled through the $S$--$D_1$ coupling. On the other hand, the dependence on the specific path $n$ can be removed. We rename the probability of the outcomes $\beta \in \{0,\pm2\}$ in Eq.~\eqref{eq:Prob_D2} as $\mathcal{P}_{\beta}(\lambda)$.

%%%%%%%%%%%%%%%%%%%%%%%%%%%%%%%%%%%%%%%%%%%%%%%%%%%%%%%%%%%%
\section{Random number generator}
With the physical setup discussed above, we construct a device that simultaneously generates random numbers and certifies the quantumness of the underlying process. A device with these features would constitute a specific semi-device-independent QRNG \cite{ma2016}, i.e., a source-independent QRNG \cite{cao2016,avesani2018}. 

Note that, here, the source of randomness is not limited to the input states, but involves the evolution of the system $S$ and its coupling with the detectors $D_1$ and $D_2$. However, our scheme does not need the user to have any knowledge of these preparations (not even of our assumptions on the state dimensions): any manipulation of the inputs would affect either the quasi-distribution obtained from $D_1$ or the observed randomness of the $D_2$ output. Moreover, no input random seed is required. Indeed, the only parameter that is tuned for the generation of the output random string is $\lambda$, whose values can be selected deterministically. On the other hand, measurements of detectors $D_1$ and $D_2$ need to be trusted.

To implement the protocol, we exploit the $2\pi$-periodicity of $\mathcal{G}_\lambda$ and discretize $[0,2\pi)$ into $L$ values $\lambda_l = 2\pi l/L$ with $l=0,1,..,L-1$.
For each value of $\lambda_l$, the experiment is repeated $K$ times.
The outcomes of $D_1$ determine $\mathrm{Re}\!\left(\mathcal{G}_{\lambda_l}\right)$ and $\mathrm{Im}\!\left(\mathcal{G}_{\lambda_l}\right)$, while $D_2$ produces a length-$K$ string $x_l=
\left\{ \beta_1^{(l)}, \beta_2^{(l)}, \dots, \beta_K^{(l)} \right\}$ with $\beta_k^{(l)} =0, \pm 2$ $\forall k,l$, see Fig.~\ref{fig:sup_path} c).
Repeating this procedure for all $l$ reconstructs $P_{ND}(\Delta)$ and generates the global length-$LK$ string $X= \{x_1,x_2,\ldots,x_{L-1}\} =
\{ \beta_1^{(1)},\beta_2^{(1)},\ldots, \beta_K^{(1)}, \beta_1^{(2)},\beta_2^{(2)},\ldots,\beta_K^{(2)}, \ldots ,\beta_K^{(L-1)} \}$.

Although the distribution $\mathcal{P}_{\beta}(\lambda_l)$ generally depends on $l$, in the large $L$ limit, its average over the complete set of couplings is $\bar{\mathcal{P}}_{\beta} \equiv \, 1/L~\sum_{l=0}^{L-1} \mathcal{P}_{\beta}(\lambda_l)=\sum_{m}' P_{m}$ because the oscillatory interference terms average to zero. 
The resulting distribution depends on both the initial state and the unitary transformation $\hat{u}$ on $S$, and, hence, can be optimized to maximize the randomness of the generated outcomes.

{\it Randomness quantification --} 
We now quantify the randomness of $X$ and relate it to the negativity of $P_{ND}$.
The quantum contribution $P_q(\Delta)$, corresponding to the interference terms with $n\neq k$, is associated with the outcomes $\Delta \in \{ \pm1,3\}$ on $D_1$ \cite{solinas2025, solinasAdvPhysX2026}. More explicitly, $P_q(-1) = 2 \re(\mathcal{A}_{4,7})$, $P_q(1) = 2 \re(\mathcal{A}_{5,8})$ and $P_q(3) = 2 \re(\mathcal{A}_{6,9})$ (see Appendix for further details). 

As expected, these quantities are completely determined by the coefficients of the initial state, $\ket{\psi_0}=\sum_i \psi_i^0\ket{i}$, and by the transition amplitudes $u_{j,i}$. As discussed above, the quantum contributions satisfy 
$P_q(-1)+P_q(1)+P_q(3)=0$,
and therefore, if they are not all zero, at least one of them must be negative \cite{solinas2025,solinasAdvPhysX2026}. 

By inspecting the relation between $P_{ND}(\Delta)$, $P_q(\Delta)$, and $\mathcal{P}_{\beta}$, we find that it determines a trade-off between certification and randomness generation.
Our goal is twofold: i) to minimize the dependence of Eq.~\eqref{eq:Prob_D2} on $\lambda$, thereby making the random-number generation process as homogeneous as possible, and ii) to make the distribution in Eq.~\eqref{eq:Prob_D2} as close to uniform as possible, thus maximizing the entropy of the generated string $X$.

The amount of randomness of the total string $X$ is quantified by the min-entropy $H_{\rm min}(X) := -\log\left[ \max_{X}{{\rm Pr}(X)} \right]$ \cite{Renyi1961,konig2009,Coles2017}.
To analyze and optimize this quantity, it is useful to decompose the probability distribution $\mathcal{P}_{\beta}$ into a classical and a quantum contribution, $\mathcal{P}_{cl}$ and $\mathcal{P}_{q}$,
corresponding to the terms that in Eq.~\eqref{eq:Prob_D2} do not depend and do depend on $\lambda$, respectively.
For the quantum contribution, explicitly, we obtain (see Appendix):
$\mathcal{P}_{q}(\beta=0) = \cos{(\lambda_l)}P_q(\Delta = -1)$, 
$\mathcal{P}_{q}(\beta=2) 
    = -\cos{(\lambda_l)}P_q(\Delta = -1)$ and 
$\mathcal{P}_{q}(\beta=-2) = 0$,
i.e., the quantum contribution is entirely determined by $P_q(-1)$ and modulates the probabilities associated with the outcomes $\beta\in\{0,\pm 2\}$ through the sampling of the quasi-characteristic function $\mathcal{G}_{\lambda_l}$.

As a result, the min-entropy of the global string $X$ is lower bounded by a function of $P_q(-1)$ only, reading (see Appendix)
\begin{equation}
    H_{\rm min}(X) \geq L K\log\left(\frac{3}{1+3|P_{q}(-1)|}\right) \equiv \bar{h}, \label{min2}
\end{equation}
with equality for $P_{q}(-1) = 0$.
Together with the certification in $D_1$, Eq.~\eqref{min2} establishes secure randomness generation against manipulations of the source (see Appendix).

This bound reveals a trade-off between certifying quantum interference through $D_1$ and maximizing the randomness generated by $D_2$. One might choose the dynamics so that $P_q(-1)=0$, which would make the output distribution independent of $\lambda_l$. However, in this scheme, $P_q(-1)=0$ also implies $P_q(1)=P_q(3)=0$ (see Appendix): all quantum contributions would then vanish, the quasiprobability distribution reconstructed from $D_1$ would become entirely positive, and the protocol would no longer certify quantum interference.

Therefore, this bound on the extractable randomness quantifies the maximum cost that our device pays in terms of randomness generation to ensure randomness certification. We can use it to upper bound the relative deviation from the maximal min-entropy - i.e., $\bar{H}_{\rm min}(X) \equiv L K \log{3}$. 
Defining $\Delta H_{\rm min}(X) \equiv |H_{\rm min}(X) - \bar{H}_{\rm min}(X)|$, the relative distance from the lower bound $\bar{h}$ in Eq.~\eqref{min2} is 
$\delta \equiv |\bar{h} - \bar{H}_{\rm min}(X)|/\bar{H}_{\rm min}(X) = \log(1+3|P_{q}(-1)|)/ \log{3} > \Delta H_{\rm min}(X)/\bar{H}_{\rm min}(X)$.
In the limit $|P_q(-1)|\ll1$, we obtain $\delta \rightarrow 3 |P_q(-1)|/\log 3$.

We see that, by engineering the negativity of the quasi-distribution (by appropriately choosing the initial state and the dynamics), we can suitably tune the bound on randomness extraction as a function of $|P_q(-1)|$.
We show in the following example that the negativity of $P_{ND}(\Delta)$ can be maximized while keeping the deviation from maximal randomness small.

%%%%%%%%%%%%%%%%%%%%%%%%%%%%%%%%%%%%%%%%%%%%
\begin{figure}
    \begin{center}
    \includegraphics[scale=.8]{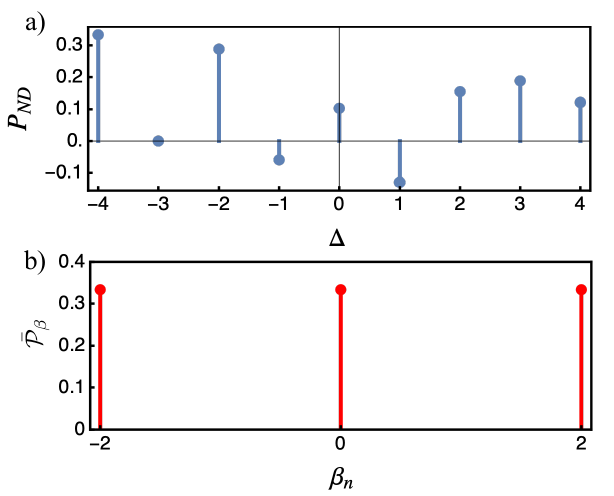}
    \end{center}
    \caption{Optimized output distributions from $D_1$ and $D_2$. Input-state parameters: $\psi_{-1}^0 \approx 0.76$, $\psi_{0}^0 \approx 0.46\,\exp{(i\phi_{0})}$, $\psi_{1}^0 \approx0.46\,\exp{(i\phi_{1})}$, $\phi_1 - \phi_0 =\pi/2$. Evolution $\hat u=\exp\!\left(i\theta\hat J_x\right)$ with $\theta = 2\arctan{\left[\sqrt{\sqrt{3}|\psi_{-1}^0|-1}\right]}\approx\pi/3$. a) Quasi-probability distribution obtained from $D_1$. The negative regions certify the presence of quantum effects.
    b) The $\lambda$-averaged distribution $\bar{\mathcal{P}}_{\beta}$ of the random numbers extracted from $D_2$. The choice of initial state and evolution allows us to obtain an almost uniformly distributed output string $X$.
    }
    \label{fig:Distributions}
\end{figure} 
%%%%%%%%%%%%%%%%%%%%%%%%%%%%%%%%%%%%%%%%%%%%
%%%%%%%%%%%%%%%%%%%%%%%%%%%%%%%%%%%%%%%%%%%%%%%%%%%%%%%%%%%%

\section{Application} 
Let us consider a specific implementation of the device described above. For a three-level system, it is convenient to use the spin-$1$ formalism. 
To fully exploit this mapping, we slightly change the notation and denote the three basis states by $\{\ket{1},\ket{0},\ket{-1}\}$, and the unitary dynamics are generated by the spin-$1$ operators $\{\hat J_x,\hat J_y,\hat J_z\}$. Within this framework, we identify $\hat a=\hat J_z$ and $\hat A=\hat J_z^2$. We take the unitary evolution to be $\hat u=\exp\!\left(i\theta\hat J_x\right)$, and the initial state 
$\ket{\psi_0}=\sum_{i=\pm1,0}\psi_i^0\ket{i}$ for the system $S$.

An example of the resulting distributions is shown in Fig.~\ref{fig:Distributions}, where we maximized $|P_q(-1)|$ over the parameters $\theta$ and $\psi_i$, and evaluated the corresponding min-entropy. Panel a) displays the quasi-probability distribution reconstructed from $D_1$. The presence of negative regions certifies the quantum nature of the process.

Panel b) shows that the $\lambda$-averaged distribution $\bar{\mathcal{P}}_{\beta}$ of the generated outcomes $\beta\in\{0,\pm2\}$ in the output string $X$ is almost uniform.
If we take into account that the randomness of the substrings $x_l\in X$ depends on $\lambda_l$, we find that the relative deviation from the completely uniform output string amounts to $\Delta H_{\rm min}(X)/\bar{H}_{\rm min}(X) \sim 9.81\%$ of the maximum extractable randomness for $L \approx 30$. In this example, we find $P_q(-1) \approx 0.06$, which fixes our upper bound on $\Delta H_{\rm min}(X)/\bar{H}_{\rm min}(X)$ to $\delta \approx 15.2\%$. We remark that this bound is not tight for $P_q(-1) \neq 0$, but it provides an estimate of the extracted randomness based on the negativity of the quasi-distribution only.

Interestingly, different choices of the initial state and/or of the evolution parameter $\theta$ can lead to different quasi-probability distributions $P_{ND}$ while yielding the same $\lambda$-averaged distribution $\bar{\mathcal{P}}_{\beta}$ of the random numbers (see Appendix). Therefore, the certification process and the statistics of the generated random numbers can be tuned independently.

%%%%%%%%%%%%%%%%%%%%%%%%%%%%%%%%%%%%%%%%%%%%%%%%%%%%%%%%%%%%
\section{Conclusions}
We have presented a novel concept for a quantum random-number generator. By exploiting three quantum systems, the device is able to generate a long string of random numbers while simultaneously certifying their quantum origin.

At its core, the protocol relies on a quantum non-demolition measurement (QNDM) scheme \cite{solinas2025,solinasAdvPhysX2026}, which reconstructs a quasi-probability distribution associated with the paths traversed during the evolution. Negative regions in this distribution signal the presence of coherent superpositions and therefore certify the genuinely quantum nature of the underlying dynamics. A second detector is used to generate the random outcomes.

The possibility of tuning both the initial state and the system dynamics opens the way to optimization strategies capable of producing a random string that is close to uniform. We showed a simple implementation where maximizing quantumness certification comes with having more than $90\%$ of the generated numbers genuinely random.

A key advantage of the protocol is that quantum certification does not require spatially separated devices. This feature opens the way to the miniaturization of the device and its integration into standard electronic platforms.

Several directions remain to be explored. Introducing additional non-demolition measurements would enable more complex dynamics and broader distributions of the generated random outcomes. Furthermore, a moderate increase in the detector dimension would provide additional degrees of freedom for optimizing both the dynamics and the measurement protocol. It is therefore plausible that, within this enlarged parameter space, further enhancements in randomness extraction can be achieved.

%%%%%%%%%%%%%%%%%%%%%%%%%%%%%%%%%%%%%%%%%%%
\section*{Acknowledgements}
The authors acknowledge fruitful discussions with Chiara Machiavello and Davide Poderini. P.S. acknowledges financial support from INFN. G.C.  acknowledges support from PNRR MUR
Project PE0000023-NQSTI.

\section*{Authors' contributions} 
P.S.~developed the idea and carried out the calculations and the numerics. G.C. performed the calculations related to randomness certification and quantification. Both authors checked the calculations, discussed the results, and wrote the manuscript. An AI-based tool was used solely for English grammar checking. The authors reviewed and approved the final text.

%%%%%%%%%%%%%%%%%%%%%%%%%%%%%%%%%%%%%%%%%%%%%%%%%%%%%%%%%%%%
%\bibliographystyle{apsrev}
%\bibliographystyle{quantum}
%\bibliography{QuasiProbabilities_QRNG}

%%%%%%%%%%%%%%%%%%%%%%%%%%%%%%%%%%%%%%%%%%%%%%%%%%%%%%%%%%%%%%%%%
%%% Appendix
%%%%%%%%%%%%%%%%%%%%%%%%%%%%%%%%%%%%%%%%%%%%%%%%%%%%%%%%%%%%%%%%%

\onecolumn
\appendix

%%%%%%%%%%%%%%%%%%%%%%%%%%%%%%%%%%%%%%%%%%%
\section{Evolution of the total system under non-demolition couplings}
\label{app:evolution}

We consider the operators $\hat a$ and $\hat A$, which act on the degrees of freedom of $S$, while $\hat\lambda$ and $\hat m$ act on $D_1$ and $D_2$, respectively. Given the coupling operators $\hat{U}_1=\exp\{ i \ha \otimes \hat\lambda \}$ (coupling $S$ to $D_1$) and $\hat{U}_2=\exp\{ \frac{2\pi i}{M} \hA \otimes \hat m \}$ (coupling $S$ to $D_2$), the operator for the non-demolition measurements is $\hU_{\rm tot} = \hU_1 \hU_2 \hu \hU_1 \hU_2$ where $\hu$ is a generic unitary operator acting on $S$.
Notice that we have used the short notation $\ha \otimes \hat \lambda \equiv \ha \otimes \hat \lambda \otimes \idop$ and $\hA \otimes \hat m \equiv \hA \otimes \idop \otimes \hat m$.

Since $\hat a$ and $\hat A$ commute, we can write the total unitary operator as 
\begin{equation}
    \hU_{\rm tot} = e^{i(\hat{a}\otimes\hat{\lambda} + \frac{2 \pi}{M} \hat{A}\otimes\hat{m})} \hat{u} e^{i(\hat{a}\otimes\hat{\lambda} + \frac{2 \pi}{M}\hat{A}\otimes\hat{m})}.
\end{equation}

The evolution of the total $S+D_1+D_2$ system for the sequential couplings is
\begin{eqnarray}
    \ket{\psi_i^0} &=& \frac{1}{\sqrt{NM}}\sum_{i,\lambda,m} \psi_i^0 \ket{i, \lambda, m} \xrightarrow{} 
    \frac{1}{\sqrt{NM}} \sum_{i,\lambda,m} \psi_i^0 e^{i \lambda a_i} e^{i \frac{2 \pi m}{M} A_i} \hat{u} \ket{i, \lambda, m}= \nonumber \\
    &=&
    \frac{1}{\sqrt{NM}} \sum_{i,j,\lambda,m} \psi_i^0 e^{i \lambda a_i} e^{i \frac{2 \pi m}{M} A_i} u_{j,i} \ket{j, \lambda, m} \nonumber \\
    &\xrightarrow{}& \frac{1}{\sqrt{NM}} \sum_{i,j,\lambda,m} \psi_i^0 e^{i \lambda (a_i+a_j)} e^{i \frac{2 \pi m}{M} (A_i+A_j)} u_{j,i} \ket{j, \lambda, m}
\end{eqnarray}
where $u_{j,i} = \bra{j} \hat{u} \ket{i}$.

The corresponding density matrix reads
\begin{eqnarray}
    \rho_f = \frac{1}{NM} \sum 
    u_{j,i} \rho^0_{i,i'} u^\dagger_{i',j'}
    e^{i \left[ \lambda (a_i+a_j)- \lambda' (a_{i'}+a_{j'}) \right]} e^{i \frac{2 \pi}{M}  \left[m (A_i+A_j) - m' (A_{i'}+A_{j'})\right] } \ketbra{j, \lambda, m}{j', \lambda', m'}
\end{eqnarray}
where the sum is over the indices $i,j,\lambda,m, i', j', \lambda', m'$. Tracing out the degrees of freedom of $S$, the density matrix of the two detectors is
\begin{eqnarray}
    \bar{\rho}_f = \frac{1}{NM} \sum 
    u_{j,i} \rho^0_{i,i'} u^\dagger_{i',j}
    e^{i \left[ \lambda (a_i+a_j)- \lambda' (a_{i'}+a_j) \right]} e^{i \frac{2 \pi}{M}  \left[m (A_i+A_j) - m' (A_{i'}+A_{j})\right] } \ketbra{ \lambda, m}{ \lambda', m'}.
    \label{app_eq:rho_D_final}
\end{eqnarray}

Each path in the eigenvalue configuration space is specified by an ordered pair ${a_i, a_j}$. Because the system has three possible eigenvalues, there are $3^2 = 9$ distinct paths in total, which we label by an integer index $n$ with $1 \leq n \leq 9$.

One convenient way to map the pair ${a_i, a_j}$ onto the index $n$ is through $n= 1+ 3 (a_i+1) +( a_j+1)$.
For instance, the pair ${1,1}$ corresponds to $n = 9$, whereas ${-1,-1}$ corresponds to $n = 1$.

With this path notation, the detectors' density matrix becomes
\begin{eqnarray}
    \bar{\rho}_f = \frac{1}{NM} \sum_{k,n=1}^9 \sum_{ \lambda, \lambda', m, m' }
    \mathcal{A}_{n,k} 
    e^{i \left( \lambda \alpha_n- \lambda' \alpha_k \right)} e^{i \frac{2 \pi}{M}  \left(m \beta_n - m' \beta_k\right) } \ketbra{ \lambda, m}{ \lambda', m'}
    \label{app_eq:rho_D_path}
\end{eqnarray}
where $\mathcal{A}_{n,k}$ is associated with the transition amplitude $u_{j,i} \rho^0_{i,i'} u^\dagger_{i',j}$ and $\alpha_n$, $\alpha_k$, $\beta_n$, $\beta_k$ are associated with $a_i+a_j$, $a_{i'}+a_j$, $A_i+A_j$, $A_{i'}+A_j$, respectively.

Notice that, with the parametrizations $n=1+3(a_i+1)+(a_j+1)$ and $k=1+3(a_{i'}+1)+(a_j+1)$, some combinations $\{n,k\}$ cannot occur.
Indeed, the constraint imposed by tracing over the final system state fixes $j$, i.e., the final state. This means that paths with different $j$ are excluded.
For example, the amplitude $\mathcal{A}_{1,2}$ is associated with the superposition of paths $\{a_1=-1,a_2=-1\}$ and $\{a_1=-1,a_2=0\}$. The final state is $\ket{-1}$ for the first path and $\ket{0}$ for the second path. Since the final state is different, there cannot be such a superposition and $\mathcal{A}_{1,2}=0$.

%%%%%%%%%%%%%%%%%%%%%%%%%%%%%%%%%%%%%%%%%%%%%%%%%%%%%%%%%%%%
\subsection{Density matrix of $D_1$}
\label{sec:rho_D1}

The density matrix of the first detector $D_1$ is obtained from Eq. \eqref{app_eq:rho_D_path} by tracing out the $D_2$ degrees of freedom.
This operation forces $m' \rightarrow m$, and we obtain
\begin{eqnarray}
    \rho_{D_1} = \frac{1}{N} \sum_{k,n=1}^9 \sum_{\lambda, \lambda'}
    \left( \frac{1}{M} \sum_{m} e^{i \frac{2 \pi m}{M}  \left( \beta_n -  \beta_k\right) } \right)
     \mathcal{A}_{n,k} 
    e^{i \left( \lambda \alpha_n- \lambda' \alpha_k \right)} \ketbra{ \lambda}{ \lambda'}.
\end{eqnarray}

The quasi-characteristic function $\mathG_\lambda$ is obtained by taking the accumulated phase on $D_1$ 
\cite{solinas2013work, solinas2015fulldistribution, solinas2016probing, solinas2025, solinasAdvPhysX2026}.
Considering that initially $D_1$ is equally weighted and taking the accumulated phase between the states $\ket{\pm \lambda}$, we have
\begin{equation}
    \mathG_\lambda  = \sum_{k,n=1}^9 
    \left( \frac{1}{M} \sum_{m} e^{i \frac{2 \pi m}{M}  \left( \beta_n -  \beta_k\right) } \right)
     \mathcal{A}_{n,k} 
    e^{i  \lambda \left(\alpha_n+ \alpha_k \right)} = \sum_{k,n=1}^9 
    f(n,k)
     \mathcal{A}_{n,k} 
    e^{i  \lambda \left(\alpha_n+ \alpha_k \right)}.
    \label{app_eq:G_lambda}
\end{equation}

For the following discussion, it is important to analyze the periodicity of $\mathG_\lambda$.
Since $\alpha_n+ \alpha_k \rightarrow 2 a_j + a_i+a_{i'}$ and, recalling that $a_i,a_{i'},a_j\in\{0,\pm1\}$, we have that $\alpha_n+ \alpha_k = 0, \pm 1, \pm 2, \pm 3, \pm 4$.
As a consequence, apart from the case with $\alpha_n+ \alpha_k =0$, the periodicity of $\mathG_\lambda$ is $2 \pi$.

The corresponding quasi-probability distribution is obtained by taking the inverse Fourier transform of $\mathG_\lambda$, so that $P_{ND} (\Delta) = 1/(2\pi) \int d \lambda \exp\{-i \lambda \Delta \}\mathG_\lambda = \sum_{k,n=1}^9 f(n,k)\mathcal{A}_{n,k}~\delta \left[\Delta - (\alpha_n+ \alpha_k) \right]$ and
\begin{equation}
    P_{ND} (\Delta) = \sum_{k,n=1}^9 
    f(n,k)
     \mathcal{A}_{n,k}~\delta \left[\Delta - (\alpha_n+ \alpha_k) \right].
     \label{app_eq:P_DN}
\end{equation}

This QNDM quasi-probability distribution has the properties discussed in Ref. \cite{solinas2025}.
The negative regions are related to the underlying presence of quantum features, i.e., coherent quantum superposition.
That is, it gives the probability distribution of the particular strings and the corresponding paths, but it also signals the presence of superposed paths during the evolution.
For example, it gives the probability for the system to go through the path/string $\{1,0\}$ but also to be in a superposition of $\{1,0\}$ and $\{-1,0\}$.
The negative regions cannot be replicated by a classical device that extracts the corresponding string from a given distribution.

Notice that $f(n,k)$ (depending on the path) acts as a decoherence functional. Indeed, if $k=n$, $f(n,n)=1$ and $\mathcal{A}_{n,n} = P_n$ is the (positive) probability to go through the $n-$th path. 
On the contrary, for $k \neq n$, $f(n,k)$ reduces the contribution of the superposed path to $ P_{ND}$.
This effect is due to the entanglement with the second detector. The trace over such degrees of freedom partially destroys the coherent evolution.

With these observations, we can rewrite the QNDM quasi-probability as 
\begin{equation}
    P_{ND} (\Delta) = P_{cl} (\Delta) + P_{q} (\Delta) = 
    \sum_{n=1}^9 
    P_n~\delta \left[\Delta - 2 \alpha_n \right] + 2 \sum_{n>k} \re \left[f(n,k)
     \mathcal{A}_{n,k} \right] 
 \delta \left[\Delta - (\alpha_n+ \alpha_k) \right].
     \label{app_eq:P_DN_simplified}
\end{equation}
where the probability of obtaining $\alpha_n$ is $P_n$.

As in Ref. \cite{solinas2025, solinasAdvPhysX2026}, it can be proven that $P_{ND} (\Delta)$ is normalized, $\sum_{n=1}^9 P_n = 1$ and $2 \sum_{n>k} \re \left[f(n,k) \mathcal{A}_{n,k} \right] =0$.
This implies that some of the off-diagonal contributions in the sum, i.e., $n \neq k$, must be negative.

%%%%%%%%%%%%%%%%%%%%%%%%%%%%%%%%%%
\subsection{Density matrix of $D_2$}
\label{sec:rho_D2}

From Eq.~\eqref{app_eq:rho_D_path}, performing the trace operation over the $D_1$ degrees of freedom, we obtain the density matrix of the second detector
\begin{eqnarray}
    \rho_{D_2} &=&  \sum_{k,n=1}^9 
    \left (\frac{1}{N} \sum_{ \lambda } e^{i  \lambda \left(\alpha_n- \alpha_k \right)}
    \right)
    \mathcal{A}_{n,k} 
    \sum_{m,m'} \frac{ e^{i \frac{2 \pi}{M}  \left(m \beta_n - m' \beta_k\right)} }{M} \ketbra{  m}{  m'} = \nonumber \\
    &=&
    \sum_{k,n=1}^9 
    F(n,k) \mathcal{A}_{n,k}  \sum_{m,m'} \frac{ e^{i \frac{2 \pi}{M}  \left(m \beta_n - m' \beta_k\right)} }{M} \ketbra{  m}{  m'}.
\end{eqnarray}

By applying the Quantum Fourier operator to $D_2$, we obtain
\begin{eqnarray}
    \rho_{D_2} =  \sum_{k,n=1}^9 
    F(n,k)~    \mathcal{A}_{n,k} 
     \ketbra{  \beta_n}{  \beta_k}.
     \label{app_eq:qftstate}
\end{eqnarray}

The probability of a measurement outcome of $D_2$ is given by a diagonal matrix element of $\rho_{D_2}$.
However, because of the degeneracy of $\hA$, different paths are associated with the same value of $\beta_n$.
Keeping this degeneracy in mind, The probability of obtaining a given outcome $\beta_n$ is therefore obtained by summing over all paths associated with the same
value $\{0,\pm2\}$. It reads
\begin{equation}
    \mathcal{P}_n (\lambda) = \sum_{m,k}' F(m,k)
    \mathcal{A}_{m,k} = \sum_{m}' P_m + 2 \sum_{m>k}' \re \left( \mathcal{A} _{m,k}\right) \cos \left[ \lambda \left(\alpha_m- \alpha_k \right) \right]
    \label{app_eq:Prob_D2}
\end{equation}
where the primed sums run over all paths corresponding to the same degenerate value $\beta_n$.

The functional $F(m,k)$ keeps track of the decoherence induced by the interaction with the detector $D_1$ and reads
\begin{equation}
    F(m,k) = \frac{1}{2} \left(
    e^{i  \lambda \left(\alpha_m - \alpha_k \right)} +
    e^{-i  \lambda \left(\alpha_m - \alpha_k \right)}
    \right) = \cos \left[ \lambda \left(\alpha_m- \alpha_k \right) \right]
\end{equation}

Equation~\eqref{app_eq:Prob_D2} gives the probability $\mathcal{P}_n$ of measuring the outcome $\beta_n$ with detector $D_2$. Since $\rho_{D_2}$ is a positive-semidefinite
density operator, its diagonal elements are nonnegative, and therefore $\mathcal{P}_n\geq0$.

%%%%%%%%%%%%%%%%%%%%%%%%%%%%%%%%%%%%%%%%%%%%%%%%
\section{Randomness quantification}
\label{app:RQSA}

Here we compute the randomness extracted from $D_2$ in terms of the min-entropy of the output random string and connect it to the randomness certification obtained from $D_1$. The min-entropy $H_{\rm min}(X)$ of a string $X$, with elements $\{x_l\}_l$ associated with a probability distribution ${\rm Pr}(X)$, is defined by the guessing probability $p_{\rm guess}(X) \equiv \max_{\{x_l\}_l}{\rm Pr}(X)$ as $H_{\rm min} \equiv -\log{p_{\rm guess}(X)}$. In our case, we have $X= \{x_l\}_{l=0}^{L-1}$ with $x_l=
\left\{ \beta_1^{(l)}, \beta_2^{(l)}, \dots, \beta_K^{(l)} \right\}$ and $\beta_k^{(l)} \in \{0, \pm 2\}$, $k=1,\ldots,K,\quad l=0,\ldots,L-1$. For any $l$ and $k$, the probability $\mathcal{P}_{\beta}$ of the outcomes $\beta_k^{(l)}$ can be decomposed into a classical $\mathcal{P}_{cl}$ and a quantum $\mathcal{P}_q$ contribution, the latter determined by superposition paths - identified by the amplitudes $\mathcal{A}_{n,k}$ with $n\neq k$. Explicitly,
\begin{align}
     \mathcal{P}_{\beta}
    &=
    \sum_{m}' P_m
    + 2
    \sum_{m>k}'
    \mathrm{Re}\!\left(\mathcal{A}_{m,k}\right)
    \cos\!\left[\lambda(\alpha_m-\alpha_k)\right] \\
    &=\sum_{m}'\mathcal{A}_{m, m} + 2
    \sum_{m>k}'
    \mathrm{Re}\!\left(\mathcal{A}_{m,k}\right)
    \cos\!\left[\lambda(\alpha_m-\alpha_k)\right] \\
    &\equiv \mathcal{P}_{cl} + \mathcal{P}_q,
    \label{app_eq:Prob_D2_decomposition}
\end{align}
where the first line is Eq.~\eqref{app_eq:Prob_D2}, in the second line we express the distributions entirely in terms of the transition amplitudes and in the last line we identify the aforementioned classical and quantum contributions. In particular, $\mathcal{P}_q$ depends on $\lambda$, the continuous parameter of the quasi-characteristic function $\mathcal{G}_{\lambda}$ evaluated in $D_1$. 
\\
Now, we briefly summarize the main operations of our device to derive the probability $p_{\rm guess}$. The discretization of $\lambda$ into $l=0,\ldots,L-1$ values $\lambda_l$ generates the $L$ substrings $x_l$ of the string $X$ and allows us to reconstruct $\mathcal{G}_{\lambda}$ by repeating the random number generation $K$ times for each $l$. Therefore, at fixed $l$, the $K$ repetitions generate the random substring $x_l$, and the 
distribution of the elements $\beta_k^{(l)}$ in $x_l$ reads $\mathcal{P}_{-2}^{k_{-2}}\mathcal{P}_{0}^{k_0}\mathcal{P}_{2}^{k_2}$, with $k_{-2} + k_0 + k_2 = K$. Before going further with the evaluation of the min-entropy, we need to derive these probabilities and connect them to the randomness certification.

\subsection{Connection between the distribution $\mathcal{P}_{\beta}$ and the quasi-distribution $P_{ND}$}
The explicit calculation of $\mathcal{P}_{\beta}$ yields
\begin{align}
    &\mathcal{P}_{-2}  = \mathcal{A}_{1,1} \\
    &\mathcal{P}_0 = \mathcal{A}_{2,2}+\mathcal{A}_{3,3}+\mathcal{A}_{4,4}+\mathcal{A}_{7,7} +2\re[\mathcal{A}_{4,7}]\cos{(\lambda_l)} \\
    &\mathcal{P}_2 = \mathcal{A}_{5,5}+\mathcal{A}_{6,6}+\mathcal{A}_{8,8}+\mathcal{A}_{9,9} +2\re[\mathcal{A}_{5,8}]\cos{(\lambda_l)}+2\re[\mathcal{A}_{6,9}]\cos{(\lambda_l)}
\end{align}
which clearly outlines the contributions $\mathcal{P}_{cl}$ and $\mathcal{P}_q$ for each $\beta$. On the other hand, by using the definition of $P_{ND}(\Delta)$, we find that the only quantum contributions $P_q(\Delta)$ are
\begin{align}   \label{negprob}
    &P_q(-1) = P_{ND}(-1) = 2\re[\mathcal{A}_{4,7}] \\
    &P_q(1) = P_{ND}(1) = 2\re[\mathcal{A}_{5,8}] \\
    &P_q(3) = P_{ND}(3) = 2\re[\mathcal{A}_{6,9}],
\end{align}
hence
\begin{align}
    &\mathcal{P}_q(\beta = 0) = \cos{(\lambda_l)}P_q(\Delta=-1) \\
    &\mathcal{P}_q(\beta = 2) = \cos{(\lambda_l)}\left[P_q(\Delta=1) + P_q(\Delta=3)\right].
\end{align}
Now we show that 
\begin{equation}
    P_q(-1) + P_q(1) + P_q(3) = 0.
\end{equation}
By expressing Eqs.~\eqref{negprob} in terms of the input state coefficients $\psi_i^0$, with $i\in\{-1,0,1\}$, and evolution $\hat{u}$, we have
\begin{align}   
    &P_q(-1) = P_{ND}(-1) = 2\re[\psi_0^0(\psi_1^0)^*\langle -1|\hat{u}|0\rangle\langle1|\hat{u}^{\dagger}|-1\rangle] \\
    &P_q(1) = P_{ND}(1) = 2\re[\psi_0^0(\psi_1^0)^*\langle 0|\hat{u}|0\rangle\langle1|\hat{u}^{\dagger}|0\rangle] \\
    &P_q(3) = P_{ND}(3) = 2\re[\psi_0^0(\psi_1^0)^*\langle 1|\hat{u}|0\rangle\langle1|\hat{u}^{\dagger}|1\rangle],
\end{align}
whose sum is reduced to
\begin{equation}
  P_q(-1) + P_q(1) + P_q(3) = \psi_0^0(\psi_1^0)^*\Trace{\left[|0\rangle\langle 1|\right]} + \psi_1^0(\psi_0^0)^*\Trace{\left[|1\rangle\langle 0|\right]} = 0. 
\end{equation}
Therefore, the quantum contribution to $\mathcal{P}_{\beta}$ depends uniquely on two quantities directly related to randomness certification, i.e., $\lambda_l$ and $P_q(-1)$. Moreover, $\mathcal{P}_q(\beta=0) = -\mathcal{P}_q(\beta=2)$.

\subsection{Randomness generation}
Finally, we need to isolate the quantum contribution $\mathcal{P}_q$ in the guessing probability in order to link randomness generation and certification. As we will see, this follows naturally if we require that the $\lambda$-averaged distribution $\bar{\mathcal{P}}_{\beta}$ is uniform, implying $\mathcal{P}_{cl}(\beta = -2) = \mathcal{P}_{cl}(\beta = 0) = \mathcal{P}_{cl}(\beta = 2) = 1/3$. In our example, where $\hat{u} = e^{i\theta\hat{J}_x}$, this constraint is satisfied by a two-parameter family of input states of the $S$ system. In general, the user does not need to check that the input states and the unitary evolutions are prepared to satisfy the condition of uniform $\mathcal{P}_{\beta}$ since any manipulation, such as replacing the randomness source with a pseudo-random-number generator, would affect the negativity of the quasi-probability distribution.

Depending on the positivity of $\cos{(\lambda_l)}$ and $P_q(-1)$, we have a definite order relation for $\mathcal{P}_q(\beta=-2)$, $\mathcal{P}_q(\beta=0)$ and $\mathcal{P}_q(\beta=2)$. In particular, the largest probability is that of $\mathcal{P}_q(\beta=0)$ and $\mathcal{P}_q(\beta=2)$ with $\cos{(\lambda_l)}P_q(-1) >0$. Since the adversary is allowed to know the input parameters and the choice of $\lambda_l$ is not randomized, he knows at each round the most probable outcome, thus limiting the actual randomness generated from $D_2$.

Therefore, for each $x_l$, the guessing probability reads 
\begin{equation} \label{guessl}
    p_{\rm guess}(x_l) = \left(\frac{1}{3} + \left|\cos{(\lambda_l)} P_{q}(-1)\right|\right)^K.
\end{equation}
Being the sets of rounds labeled by $l$ independent and each one characterized by the guessing probability in Eq.~\eqref{guessl}, $p_{\rm guess}(X)$ factorizes and the min-entropy of the global string $X$ reads

\begin{equation}
    H_{\rm min}(X) = -K\sum_{l=0}^{L-1} \log\left(\frac{1}{3} + \left|\cos{(\lambda_l)} P_{q}(-1)\right|\right), \label{min}
\end{equation}
and one can easily check that it is bounded from below by Eq.~$(5)$ of the main text.

\section{Application}
\label{app:app}
Given the evolution $\hat{u} = e^{i\theta\hat{J}_x}$ considered in our letter, the condition $\mathcal{P}_{cl}(\beta = -2) = \mathcal{P}_{cl}(\beta = 0) = \mathcal{P}_{cl}(\beta = 2) = 1/3$ constrains the coefficients $\psi_i^0 = |\psi_i|e^{i\phi_i}$ of the input state of system $S$ and the coupling $\theta$ of the evolution $\hat{u}$ as follows,
\begin{align} \label{par1}
    &|\psi_0|^2 = \frac{1-2\sqrt{3}|\psi_{-1}|(1-|\psi_{-1}|^2)}{(\sqrt{3}|\psi_{-1}|-1)(\sqrt{3}|\psi_{-1}|-3)} \\
    &|\psi_{1}|^2 = \frac{2(1-\sqrt{3}|\psi_{-1}|)+\sqrt{3}|\psi_{-1}|^3(2-\sqrt{3}|\psi_{-1}|)}{(\sqrt{3}|\psi_{-1}|-1)(\sqrt{3}|\psi_{-1}|-3)} \label{par2} \\
    &\theta = \pm2\arctan{\left(\sqrt{\sqrt{3}|\psi_{-1}|-1}\right)} + 2c\pi \label{par3}
\end{align}
with $c \in \mathbb{Z}$.

The quantum contributions $P_q(\Delta)$ of the quasi-probability distribution $P_{ND}$ read
\begin{align}
&P_q(-1) = \frac{1}{\sqrt{2}}\sin{(\phi_1-\phi_0)}|\psi_1||\psi_0|\sin{\theta}(1-\cos{\theta}) \label{par4} \\
&P_q(1) = \frac{1}{\sqrt{2}}\sin{(\phi_1-\phi_0)}|\psi_1||\psi_0|\sin{(2\theta)} \label{par5} \\
&P_q(3) = -\frac{1}{\sqrt{2}}\sin{(\phi_1-\phi_0)}|\psi_1||\psi_0|\sin{\theta}(1+\cos{\theta}). \label{par6}
\end{align}
Therefore, through Eqs.~\eqref{par1},~\eqref{par2} and~\eqref{par3}, they can all be expressed in terms of $|\psi_{-1}|$ and $\Delta\phi\equiv\phi_1-\phi_0$, i.e., they are functions of two parameters only, which allows us to continuously tune the negativity of $P_{ND}$ - and hence the randomness of the output string $X$, as shown in Eq.~\eqref{min}. In our application, we maximize $|P_q(-1)|$ over $\psi_{-1}$ and $\Delta\phi$ and show that, despite having a trade-off between randomness certification and generation, more than $90\%$ of the generated numbers are uniformly distributed with optimal certification. 

%%%%%%%%%%%%%%%%%%%%%%%%%%%%%%%%%%%%%%%%%%%%%%%%%%%%%%%%%%%%
\bibliographystyle{quantum}
\bibliography{QuasiProbabilities_QRNG}

%%%%%%%%%%%%%%%%%%%%%%%%%%%%%%%%%%%%%%%%%%%%%%%%%%%%%%%%%%%%

%%%%%%%%%%%%%%%%%%%%%%%%%%%%%%%%%%%%%%%%%%%%%%%%%%%%%%%%%%%%%%

%%%%%%%%%%%%%%%%%%%%%%%%%%%%%%%%%%%%%%%%%%%%%%%%%%%%%%%%%%%%

 \end{document}